# Octonionic Strong and Weak Interactions and Their Quantum Equations

Zihua Weng
*School of Physics and Mechanical & Electrical Engineering, P. O. Box 310, Xiamen University, Xiamen 361005, China*

**Abstract**

By analogy with the octonionic electromagnetic and gravitational interactions, the octonionic strong and weak interactions and their quantum interplays are discussed in the paper. In the weak interaction, the study deduces some conclusions of field sources and intermediate particles, which are consistent with the Dirac equation, Yang-Mills equation and some new intermediate particles etc. In the strong interaction, the research draws some conclusions of the field sources and intermediate particles, which are coincident with the Dirac-like equation, three kinds of colors and some new intermediate particles etc. The researches results show that there exist some new field source particles in the strong and weak interactions.

*Keywords*: octonion space; strong interaction; weak interaction; quantum; color.

## 1. Introduction

The physics of strong and weak interactions is undoubtedly one of the most challenging areas of modern science. The strong and weak interactions keep providing new experimental observations, which were not predicted by 'effective' theories. It retains the problems in describing all the observed phenomena simultaneously. The deep understanding might have further far-reaching implications in strong and weak interactions.

A new insight on the problem of strong and weak interactions can be given by the concept of the octonionic space. By analogy with the electromagnetic-gravitational field theory, the concepts of the octonionic space and four kinds of subfields can be spread to the strong-weak filed. According to previous research results of the electromagnetic-gravitational field and the 'SpaceTime Equality Postulation', the strong and weak interactions can be described by the quaternionic space [1]. Based on the conception of the space verticality etc., two types of quaternionic spaces can be united into an octonionic space. In the octonionic space, the strong and weak interactions can be equally described. So the physical characteristics of field sources particles and intermediate particles of the strong-strong, weak-weak, strong-weak and weak- strong subfields can be described by the octonionic space uniformly.

---

*E-mail Addresses*: xmuwzh@hotmail.com, xmuwzh@xmu.edu.cn

The paper describes the strong-weak field and their quantum theory, and draws some new conclusions which are consistent with the Yang-Mills equation and three kinds of colors etc. In the octonionic strong-weak field, a few predictions which are associated with the quantum movement feature of quarks and leptons can be deduced, and some unknown and new intermediate particles can be achieved in the study.

## 2. Octonionic electromagnetic-gravitational field

The electromagnetic and gravitational interactions are interconnected, unified and equal. Both of them can be described in the quaternionic space. By means of the conception of the space expansion etc., two types of quaternionic spaces can combine into an octonionic space (Octonionic space E-G, for short). In the octonionic space, various characteristics of electromagnetic and gravitational interactions can be described uniformly, and some equations set of the electromagnetic-gravitational field can be attained.

There exist four types of subfields and their field sources of electromagnetic-gravitational field. In Table 1, the electromagnetic-electromagnetic subfield is 'extended electromagnetic field', and its general charge is E electric-charge. The gravitational-gravitational subfield is 'modified gravitational filed', and its general charge is G gravitational-charge. Meanwhile, electromagnetic-gravitational and gravitational-electromagnetic subfields are both long range fields and candidates of the 'dark matter field'. And their general charges (G electric-charge and E gravitational-charge) are candidates of 'dark matter'. The physical features of the dark matter meet the requirement of field equations set in Table 2.

Table 1. Subfield types of electromagnetic-gravitational field

| | Electromagnetic Interaction | Gravitational Interaction |
|---|---|---|
| electromagnetic quaternionic space | electromagnetic-electromagnetic subfield, E electric-charge, intermediate particle $\gamma_{ee}$ | gravitational-electromagnetic subfield, E gravitational-charge, intermediate particle $\gamma_{ge}$ |
| gravitational quaternionic space | electromagnetic-gravitational subfield, G electric-charge, intermediate particle $\gamma_{eg}$ | gravitational-gravitational subfield, G gravitational-charge, intermediate particle $\gamma_{gg}$ |

The electromagnetic-gravitational and gravitational-electromagnetic subfields are both long range fields and candidates of dark matter field. The field strength of the electromagnetic-gravitational and gravitational-electromagnetic subfields may be equal and a little less than that of gravitational-gravitational subfield. The dark matter field has the following prediction. Two types of field sources possessed by the dark matter field would make the dark matter particles diversiform. The research results explain that some observed abnormal phenomena about celestial bodies are caused by either modified gravitational interaction or dark matter.

Table 2. Equations set of electromagnetic-gravitational field

| Spacetime | Octonionic space E-G |
|---|---|
| $\mathcal{X}$ physical quantity | $\mathcal{X} = \mathcal{X}_{E\text{-}G}$ |
| Field potential | $\mathcal{A} = \lozenge^* \circ \mathcal{X}$ |
| Field strength | $\mathcal{B} = \lozenge \circ \mathcal{A}$ |
| Field source | $c\mu\mathcal{S} = c\,(\mathcal{B}/c + \lozenge)^* \circ \mathcal{B}$ |
| Force | $\mathcal{Z} = c\,(\mathcal{B}/c + \lozenge) \circ \mathcal{S}$ |
| Angular momentum | $\mathcal{M} = \mathcal{S} \circ (\mathcal{R} + k_{rx}\mathcal{X})$ |
| Energy | $\mathcal{W} = c\,(\mathcal{B}/c + \lozenge)^* \circ \mathcal{M}$ |
| Power | $\mathcal{N} = c\,(\mathcal{B}/c + \lozenge) \circ \mathcal{W}$ |

In the octonionic space E-G, the wave functions of the quantum mechanics are the octonionic equations set. And the Dirac and Klein-Gordon equations of the quantum mechanics are actually the wave equations set which are associated with particle's wave function $\Psi = \mathcal{M}/\hbar$. Wherein, the wave function $\Psi$ can be written as the exponential form with the octonionic characteristics. The coefficient h is Planck constant, and $\hbar = h/2\pi$.

By comparison, we find that the Dirac and Klein-Gordon equations can be attained respectively from the energy quantum equation and the power quantum equations after substituting the operator $c\,(\mathcal{B}/c + \lozenge)$ for $\{\mathcal{W}/(c\hbar) + \lozenge\}$. By analogy with the above equation, the Dirac-like and the second Dirac-like equations can be procured from the field source and force quantum equations respectively. Wherein, $\mathcal{B}/\hbar$ is the wave function also. [2]

Table 3. Quantum equations set of electromagnetic-gravitational field

| | |
|---|---|
| Energy quantum | $\mathcal{U} = (\mathcal{W}/c + \hbar\lozenge)^* \circ (\mathcal{M}/\hbar)$ |
| Power quantum | $\mathcal{L} = (\mathcal{W}/c + \hbar\lozenge) \circ (\mathcal{U}/\hbar)$ |
| Field source quantum | $\mathcal{T} = (\mathcal{W}/c + \hbar\lozenge)^* \circ (\mathcal{B}/\hbar)$ |
| Force quantum | $\mathcal{O} = (\mathcal{W}/c + \hbar\lozenge) \circ (\mathcal{T}/\hbar)$ |

## 3. Octonionic strong-weak field

### 3.1 Octonionic space

By analogy with the octonionic spacetime of electromagnetic-gravitational field, the strong-weak field possesses its own octonionic spacetime. According to the 'SpaceTime Equality Postulation', the spacetime, which is associated with the strong interaction and possesses the physics content, is adopted by the quaternionic space. And the spacetime derived from the weak interaction is supposed to be the quaternionic space also.

The strong and weak interactions are interconnected, unified and equal. Both of them can be described in the quaternionic space. By means of the conception of the space expansion etc., two types of the quaternionic spaces can combine into an octonionic space (Octonionic space S-W, for short). In the octonionic space, various characteristics of the strong and weak interactions can be described uniformly, and some equations set of the strong-weak field can be obtained.

The base $\mathcal{E}_s$ of the quaternionic space (S space, for short) of the strong interaction is $\mathcal{E}_s$ = ($\mathbf{1}$, $\vec{i}$ , $\vec{j}$ , $\vec{k}$ ) . The base $\mathcal{E}_w$ of the quaternionic space (W space, for short) of the weak interaction is independent of the base $\mathcal{E}_s$. Selecting $\mathcal{E}_w = (\mathbf{1}, \vec{i}, \vec{j}, \vec{k}) \circ \vec{e} = (\vec{e}, \vec{I}, \vec{J}, \vec{K})$. So the base $\mathcal{E}_s$ and $\mathcal{E}_w$ can constitute the base $\mathcal{E}$ of the octonionic space S-W.

$$\mathcal{E} = \mathcal{E}_s + \mathcal{E}_w = (\mathbf{1}, \vec{i}, \vec{j}, \vec{k}, \vec{e}, \vec{I}, \vec{J}, \vec{K}) \qquad (1)$$

Table 4. Octonion multiplication table

| | $\mathbf{1}$ | $\vec{i}$ | $\vec{j}$ | $\vec{k}$ | $\vec{e}$ | $\vec{I}$ | $\vec{J}$ | $\vec{K}$ |
|---|---|---|---|---|---|---|---|---|
| $\mathbf{1}$ | $\mathbf{1}$ | $\vec{i}$ | $\vec{j}$ | $\vec{k}$ | $\vec{e}$ | $\vec{I}$ | $\vec{J}$ | $\vec{K}$ |
| $\vec{i}$ | $\vec{i}$ | **-1** | $\vec{k}$ | $-\vec{j}$ | $\vec{I}$ | $-\vec{e}$ | $-\vec{K}$ | $\vec{J}$ |
| $\vec{j}$ | $\vec{j}$ | $-\vec{k}$ | **-1** | $\vec{i}$ | $\vec{J}$ | $\vec{K}$ | $-\vec{e}$ | $-\vec{I}$ |
| $\vec{k}$ | $\vec{k}$ | $\vec{j}$ | $-\vec{i}$ | **-1** | $\vec{K}$ | $-\vec{J}$ | $\vec{I}$ | $-\vec{e}$ |
| $\vec{e}$ | $\vec{e}$ | $-\vec{I}$ | $-\vec{J}$ | $-\vec{K}$ | **-1** | $\vec{i}$ | $\vec{j}$ | $\vec{k}$ |
| $\vec{I}$ | $\vec{I}$ | $\vec{e}$ | $-\vec{K}$ | $\vec{J}$ | $-\vec{i}$ | **-1** | $-\vec{k}$ | $\vec{j}$ |
| $\vec{J}$ | $\vec{J}$ | $\vec{K}$ | $\vec{e}$ | $-\vec{I}$ | $-\vec{j}$ | $\vec{k}$ | **-1** | $-\vec{i}$ |
| $\vec{K}$ | $\vec{K}$ | $-\vec{J}$ | $\vec{I}$ | $\vec{e}$ | $-\vec{k}$ | $-\vec{j}$ | $\vec{i}$ | **-1** |

The displacement ( $r_0$ , $r_1$ , $r_2$ , $r_3$ , $R_0$ , $R_1$ , $R_2$ , $R_3$ ) in the octonionic space is

$$\mathcal{R} = ( r_0 + \vec{i}\, r_1 + \vec{j}\, r_2 + \vec{k}\, r_3) + ( \vec{e}\, R_0 + \vec{I}\, R_1 + \vec{J} R_2 + \vec{K}\, R_3) \qquad (2)$$

where, $r_0$ = ct, $R_0$ = cT. c is the speed of intermediate particles, t and T denote the time.

The octonionic differential operator $\Diamond$ and its conjugate operator $\Diamond^*$ are defined as,

$$\Diamond = \Diamond_s + \Diamond_w \quad , \quad \Diamond^* = \Diamond^*_s + \Diamond^*_w \qquad (3)$$

where, $\Diamond_s = \partial_{s0} + \vec{i}\, \partial_{s1} + \vec{j}\, \partial_{s2} + \vec{k}\, \partial_{s3}$ , $\Diamond_w = \vec{e}\, \partial_{w0} + \vec{I}\, \partial_{w1} + \vec{J} \partial_{w2} + \vec{K}\, \partial_{w3}$ . $\partial_{sj} = \partial / \partial\, r_j$ , $\partial_{wj} = \partial / \partial\, R_j$ , j = 0, 1, 2, 3 .

After checking, the octonionic differential operator meets ($\mathcal{Q}$ is an octonion)

$$\Diamond^* \circ (\Diamond \circ \mathcal{Q}) = (\Diamond^* \circ \Diamond) \circ \mathcal{Q} = (\Diamond \circ \Diamond^*) \circ \mathcal{Q} \qquad (4)$$

In the strong-weak field, the field potential ( $a_0$ , $a_1$ , $a_2$ , $a_3$ , $k_{sw} A_0$ , $k_{sw} A_1$ , $k_{sw} A_2$ , $k_{sw} A_3$ ) is defined as

$$\mathcal{A} = \Diamond^* \circ \mathcal{X}$$
$$= (a_0 + \vec{i}\, a_1 + \vec{j}\, a_2 + \vec{k}\, a_3) + k_{sw} ( \vec{e}\, A_0 + \vec{I}\, A_1 + \vec{J} A_2 + \vec{K}\, A_3) \qquad (5)$$

where, $k_{rx}\, \mathcal{X} = k^s_{rx}\, \mathcal{X}_s + k^w_{rx}\, \mathcal{X}_w$ ; $\mathcal{X}_s$ is the physical quantity in S space, and $\mathcal{X}_w$ is the physical quantity in W space; $k_{sw}$ is the coefficient.

### 3.2 Equations set of strong-weak field

In strong-weak field, there exist four types of subfields and their field sources. The subfields of the strong-weak field are short range. In Table 5, the strong-strong subfield may be 'extended strong field', its general charge is S strong-charge. The weak-weak subfield may be regarded as the 'modified weak field', and its general charge is W weak-charge. And the strong-weak and weak-strong subfields may be regarded as two kinds of new and unknown subfields. Their general charges are W strong-charge and S weak-charge respectively.

It can be predicted that the field strength of the strong-weak and weak-strong subfields may be equal, and both of them are weaker than that of weak-weak subfield. Sometimes, the first two subfields may be regarded as 'familiar weak field' mistakenly. Therefore three kinds of colors have to introduce into the familiar weak interaction theory to distinguish above three types of weaker subfields of strong-weak field. The physical features of each subfield of the strong-weak field meet the requirement of Eqs.(6) ~ (11).

Table 5. Subfield types of strong-weak field

| | Strong Interaction | Weak Interaction |
|---|---|---|
| strong quaternionic space | strong-strong subfield<br>S strong-charge<br>Intermediate particle $\gamma_{ss}$ | weak-strong subfield<br>S weak-charge<br>intermediate particle $\gamma_{ws}$ |
| weak quaternionic space | strong-weak subfield<br>W strong-charge<br>Intermediate particle $\gamma_{sw}$ | weak-weak subfield<br>W weak-charge<br>intermediate particle $\gamma_{ww}$ |

In the strong-weak field, the definition of the field strength is very different from that of the electromagnetic-gravitational field, and thus the succeeding equations and operators should be revised. By analogy with the case of the electromagnetic-gravitational field, the octonionic differential operator $\Diamond$ in the strong-weak field needs to be generalized to the new operator $(\mathcal{A}/k + \Diamond)$. That is because of the field potential is much more fundamental than the field strength, and the strong-weak field belongs to short range. Therefore physical characteristics of the strong-weak field can be researched from many aspects. Where, k = c is the constant of the strong-weak field.

The field strength $\mathcal{B}$ of the strong-weak field can be defined as

$$\mathcal{B} = (\mathcal{A}/k + \Diamond) \circ \mathcal{A} \tag{6}$$

then the field source and the force of the strong-weak field can be defined respectively as

$$k\mu \mathcal{S} = k\,(\mathcal{A}/k + \Diamond)^* \circ \mathcal{B} \tag{7}$$

$$\mathcal{Z} = k\,(\mathcal{A}/k + \Diamond) \circ \mathcal{S} \tag{8}$$

where, the mark (*) denotes octonionic conjugate. The coefficient μ is interaction intensity of the strong-weak field.

As a part of the field source $\mathcal{S}$, the item $(\mathcal{A}^* \circ \mathcal{B}/k)$ has an important impact on the subsequent equations. In strong-weak field, the conserved zero-field can be attained when the item $\Diamond^* \circ \mathcal{B} = 0$, the force-balance equation can be achieved when the force $\mathcal{Z} = 0$.

The angular momentum of the strong-weak field can be defined as

$$\mathcal{M} = \mathcal{S} \circ (\mathcal{R} + k_{rx} \mathcal{X}) \tag{9}$$

and the energy and power in the strong-weak field can be defined respectively as

$$\mathcal{W} = k\,(\mathcal{A}/k + \diamondsuit)^* \circ \mathcal{M} \tag{10}$$

$$\mathcal{N} = k\,(\mathcal{A}/k + \diamondsuit) \circ \mathcal{W} \tag{11}$$

The above equations show that, the angular momentum $\mathcal{M}$, energy $\mathcal{W}$ and power $\mathcal{N}$ of the strong-weak field are different to that of the electromagnetic-gravitational field. The physical quantity $\mathcal{X}$ has effect on the field potential $\diamondsuit^* \circ \mathcal{X}$, the angular momentum $\mathcal{S} \circ \mathcal{X}$, the energy $\mathcal{A}^* \circ (\mathcal{S} \circ \mathcal{X})$ and the power $(\mathcal{A} \circ \mathcal{A}^*) \circ (\mathcal{S} \circ \mathcal{X})$. The introduction of the physical quantity $\mathcal{X}$ makes the definition of the angular momentum $\mathcal{M}$ and the energy $\mathcal{W}$ more integrated and complicated, and the theory more self-consistent.

In the above equations, the conservation of the angular momentum in the strong-weak field can be gained when $\mathcal{W} = 0$, and the energy conservation of the strong-weak field can be attained when $\mathcal{N} = 0$.

Table 6. Equations set of strong-weak field

| Spacetime | Octonionic space S-W |
|---|---|
| $\mathcal{X}$ physical quantity | $\mathcal{X} = \mathcal{X}_{S\text{-}W}$ |
| Field potential | $\mathcal{A} = \diamondsuit^* \circ \mathcal{X}$ |
| Field strength | $\mathcal{B} = (\mathcal{A}/k + \diamondsuit) \circ \mathcal{A}$ |
| Field source | $k\mu \mathcal{S} = k\,(\mathcal{A}/k + \diamondsuit)^* \circ \mathcal{B}$ |
| Force | $\mathcal{Z} = k\,(\mathcal{A}/k + \diamondsuit) \circ \mathcal{S}$ |
| Angular momentum | $\mathcal{M} = \mathcal{S} \circ (\mathcal{R} + k_{rx} \mathcal{X})$ |
| Energy | $\mathcal{W} = k\,(\mathcal{A}/k + \diamondsuit)^* \circ \mathcal{M}$ |
| Power | $\mathcal{N} = k\,(\mathcal{A}/k + \diamondsuit) \circ \mathcal{W}$ |

## 4. Quantum equations set of strong-weak field

In the octonionic space S-W, the wave functions of the quantum mechanics are the octonionic equations set. And the Dirac and Klein-Gordon equations of the quantum mechanics are actually the wave equations set which are associated with particle's wave function $\Psi = \mathcal{M}/b$. Wherein, the wave function $\Psi = \mathcal{S} \circ (\mathcal{R} + k_{rx} \mathcal{X})/b$ can be written as the exponential form with the octonionic characteristics. The coefficient $b = \hbar_{S\text{-}W}$ is the Plank-like constant.

### 4.1 Equations set of Dirac and Klein-Gordon

By comparison, we find that the Dirac equation and Klein-Gordon equation can be attained respectively from the energy equation (10) and power equation (11) after substituting the operator $k\,(\mathcal{A}/k + \diamondsuit)$ for $(\mathcal{W}/cb + \diamondsuit)$.

The $\mathcal{U}$ equation of the quantum mechanics can be defined as

$$\mathcal{U} = (\mathcal{W}/c + b\diamondsuit)^* \circ (\mathcal{M} / b) \tag{12}$$

The $\mathcal{L}$ equation of the quantum mechanics can be defined as

$$\mathcal{L} = (\mathcal{W}/c + b\Diamond) \circ (\mathcal{U} / b) \qquad (13)$$

From the energy quantum equation $\mathcal{U}$ = 0, the Dirac and Schrodinger equations in the strong-weak field can be deduced to describe the field source particle with spin 1/2 (quark and lepton etc.). From the power quantum equation $\mathcal{L}$ = 0, the Klein-Gordon equation in the strong-weak field can be obtained to explain the field source particle with spin 0.

### 4.2 Equations set of Dirac-like

By analogy with the above equations, three sorts of Dirac-like equations can be procured from the field strength equation (6), field source equation (7) and force equation (8) respectively. Wherein, $\mathcal{A}$ /b is the wave function also.

The $\mathcal{G}$ equation of quantum mechanics can be defined as

$$\mathcal{G} = (\mathcal{W}/c + b\Diamond) \circ (\mathcal{A} / b) \qquad (14)$$

The $\mathcal{T}$ equation of quantum mechanics can be defined as

$$\mathcal{T} = (\mathcal{W}/c + b\Diamond)^* \circ (\mathcal{G} / b) \qquad (15)$$

The $\mathcal{O}$ equation of quantum mechanics can be defined as

$$\mathcal{O} = (\mathcal{W}/c + b\Diamond) \circ (\mathcal{T} / b) \qquad (16)$$

From the field strength quantum equation $\mathcal{G}$ = 0, Eq.(14) in the strong-weak field can be achieved to make out the first type of intermediate particles (intermediate boson etc). From the field source quantum equation $\mathcal{T}$ = 0, Eq.(15) in the strong-weak field can be described the second type of intermediate particles. From the force quantum equation $\mathcal{O}$ = 0, Eq.(16) in the strong-weak field can be attained to explain the intermediate particles.

Table 7. Quantum equations set of strong-weak field

| | |
|---|---|
| Energy quantum | $\mathcal{U} = (\mathcal{W}/c + b\Diamond)^* \circ (\mathcal{M} /b)$ |
| Power quantum | $\mathcal{L} = (\mathcal{W}/c + b\Diamond) \circ (\mathcal{U} /b)$ |
| Field strength quantum | $\mathcal{G} = (\mathcal{W}/c + b\Diamond) \circ (\mathcal{A} /b)$ |
| Field source quantum | $\mathcal{T} = (\mathcal{W}/c + b\Diamond)^* \circ (\mathcal{G} /b)$ |
| Force quantum | $\mathcal{O} = (\mathcal{W}/c + b\Diamond) \circ (\mathcal{T} /b)$ |

## 5. Approximate cases

In the Eq.(6) of strong-weak field, the Yang-Mills equation can be achieved when the field potential are limited to the components ($a_0$ , $a_1$ , $a_2$ , $a_3$ ). In the Table 5, each flavor of quark will occupy three kinds of colors to distinguish different quarks, when the quark particles possess the S strong-charge together with other general charges. In strong-weak field, Dirac equation and Klein-Gordon equations can be gained from Eqs.(12) and (13) respectively. The Dirac-like equation can be obtained from Eq.(14), and other sorts of equations set can be attained from the Eqs.(15) and (16) respectively.

### 5.1 Yang-Mills equations

In the octonionic space, the field potential (5) of strong-weak field can be rewritten as

$$\mathcal{A} = a_0 + \vec{i} \circ a_1 + \vec{j} \circ a_2 + \vec{k} \circ a_3 + \vec{e} \circ \mathcal{A}_0 + \vec{I} \circ \mathcal{A}_1 + \vec{J} \circ \mathcal{A}_2 + \vec{K} \circ \mathcal{A}_3$$

where, the field potential components $a_j$ and $\mathcal{A}_j$ are the octonionic function.

In the strong-weak field, the field strength equation (6) can be decomposed to

$$\begin{aligned}
\mathcal{B} &= (\mathcal{A}/\mathrm{k} + \lozenge) \circ \mathcal{A} \\
&= \{ (\partial_{s0} + \vec{i}\,\partial_{s1} + \vec{j}\,\partial_{s2} + \vec{k}\,\partial_{s3} + \vec{e}\,\partial_{w0} + \vec{I}\,\partial_{w1} + \vec{J}\,\partial_{w2} + \vec{K}\,\partial_{w3}) \\
&\qquad + (a_0 + \vec{i} \circ a_1 + \vec{j} \circ a_2 + \vec{k} \circ a_3 + \vec{e} \circ \mathcal{A}_0 + \vec{I} \circ \mathcal{A}_1 + \vec{J} \circ \mathcal{A}_2 + \vec{K} \circ \mathcal{A}_3)/\mathrm{k} \, \} \\
&\qquad \circ (a_0 + \vec{i} \circ a_1 + \vec{j} \circ a_2 + \vec{k} \circ a_3 + \vec{e} \circ \mathcal{A}_0 + \vec{I} \circ \mathcal{A}_1 + \vec{J} \circ \mathcal{A}_2 + \vec{K} \circ \mathcal{A}_3) \\
&= \mathcal{B}^1{}_{00} + \mathcal{B}^1{}_{01} + \mathcal{B}^1{}_{02} + \mathcal{B}^1{}_{03} + \mathcal{B}^1{}_{23} + \mathcal{B}^1{}_{31} + \mathcal{B}^1{}_{12} \\
&\quad + \mathcal{B}^2{}_{00} + \mathcal{B}^2{}_{01} + \mathcal{B}^2{}_{02} + \mathcal{B}^2{}_{03} + \mathcal{B}^2{}_{23} + \mathcal{B}^2{}_{31} + \mathcal{B}^2{}_{12} \\
&\quad + \mathcal{B}^3{}_{00} + \mathcal{B}^3{}_{01} + \mathcal{B}^3{}_{02} + \mathcal{B}^3{}_{03} + \mathcal{B}^3{}_{23} + \mathcal{B}^3{}_{31} + \mathcal{B}^3{}_{12} \\
&\quad + \mathcal{B}^4{}_{00} + \mathcal{B}^4{}_{01} + \mathcal{B}^4{}_{02} + \mathcal{B}^4{}_{03} + \mathcal{B}^4{}_{23} + \mathcal{B}^4{}_{31} + \mathcal{B}^4{}_{12}
\end{aligned} \tag{17}$$

where, the field strength can be separated to the following octonionic components,

$$\begin{aligned}
\mathcal{B}^1{}_{00} &= \{ \partial_{s0}\, a_0 - \partial_{s1}\, a_1 - \partial_{s2}\, a_2 - \partial_{s3}\, a_3 \} \\
&\quad + \{ a_0 \circ a_0 + (\vec{i} \circ a_1) \circ (\vec{i} \circ a_1) + (\vec{j} \circ a_2) \circ (\vec{j} \circ a_2) + (\vec{k} \circ a_3) \circ (\vec{k} \circ a_3) \}/\mathrm{k} \\
\mathcal{B}^1{}_{01} &= \vec{i} \circ (\partial_{s0}\, a_1 + \partial_{s1}\, a_0) + \{ a_0 \circ (\vec{i} \circ a_1) + (\vec{i} \circ a_1) \circ a_0 \}/\mathrm{k} \\
\mathcal{B}^1{}_{02} &= \vec{j} \circ (\partial_{s0}\, a_2 + \partial_{s2}\, a_0) + \{ a_0 \circ (\vec{j} \circ a_2) + (\vec{j} \circ a_2) \circ a_0 \}/\mathrm{k} \\
\mathcal{B}^1{}_{03} &= \vec{k} \circ (\partial_{s0}\, a_3 + \partial_{s3}\, a_0) + \{ a_0 \circ (\vec{k} \circ a_3) + (\vec{k} \circ a_3) \circ a_0 \}/\mathrm{k} \\
\mathcal{B}^1{}_{23} &= \{ \vec{j} \circ (\vec{k} \circ \partial_{s2}\, a_3) + \vec{k} \circ (\vec{j} \circ \partial_{s3}\, a_2) \} \\
&\quad + \{ (\vec{j} \circ a_2) \circ (\vec{k} \circ a_3) + (\vec{k} \circ a_3) \circ (\vec{j} \circ a_2) \}/\mathrm{k} \\
\mathcal{B}^1{}_{31} &= \{ \vec{k} \circ (\vec{i} \circ \partial_{s3}\, a_1) + \vec{i} \circ (\vec{k} \circ \partial_{s1}\, a_3) \} \\
&\quad + \{ (\vec{k} \circ a_3) \circ (\vec{i} \circ a_1) + (\vec{i} \circ a_1) \circ (\vec{k} \circ a_3) \}/\mathrm{k} \\
\mathcal{B}^1{}_{12} &= \{ \vec{i} \circ (\vec{j} \circ \partial_{s1}\, a_2) + \vec{j} \circ (\vec{i} \circ \partial_{s2}\, a_1) \} \\
&\quad + \{ (\vec{i} \circ a_1) \circ (\vec{j} \circ a_2) + (\vec{j} \circ a_2) \circ (\vec{i} \circ a_1) \}/\mathrm{k} \\
\mathcal{B}^2{}_{00} &= \{ \vec{e} \circ \partial_{w0}\, a_0 + \vec{I} \circ (\vec{i} \circ \partial_{w1}\, a_1) + \vec{J} \circ (\vec{j} \circ \partial_{w2}\, a_2) + \vec{K} \circ (\vec{k} \circ \partial_{w3}\, a_3) \} \\
&\quad + \{ a_0 \circ (\vec{e} \circ \mathcal{A}_0) + (\vec{i} \circ a_1) \circ (\vec{I} \circ \mathcal{A}_1) + (\vec{j} \circ a_2) \circ (\vec{J} \circ \mathcal{A}_2) + (\vec{k} \circ a_3) \circ (\vec{K} \circ \mathcal{A}_3) \}/\mathrm{k} \\
\mathcal{B}^2{}_{01} &= \{ \vec{e} \circ (\vec{i} \circ \partial_{w0}\, a_1) + \vec{I} \circ \partial_{w1}\, a_0 \} + \{ a_0 \circ (\vec{I} \circ \mathcal{A}_1) + (\vec{i} \circ a_1) \circ (\vec{e} \circ \mathcal{A}_0) \}/\mathrm{k} \\
\mathcal{B}^2{}_{02} &= \{ \vec{e} \circ (\vec{j} \circ \partial_{w0}\, a_2) + \vec{J} \circ \partial_{w2}\, a_0 \} + \{ a_0 \circ (\vec{J} \circ \mathcal{A}_2) + (\vec{j} \circ a_2) \circ (\vec{e} \circ \mathcal{A}_0) \}/\mathrm{k} \\
\mathcal{B}^2{}_{03} &= \{ \vec{e} \circ (\vec{k} \circ \partial_{w0}\, a_3) + \vec{K} \circ \partial_{w3}\, a_0 \} + \{ a_0 \circ (\vec{K} \circ \mathcal{A}_3) + (\vec{k} \circ a_3) \circ (\vec{e} \circ \mathcal{A}_0) \}/\mathrm{k} \\
\mathcal{B}^2{}_{23} &= \{ \vec{J} \circ (\vec{k} \circ \partial_{w2}\, a_3) + \vec{K} \circ (\vec{j} \circ \partial_{w3}\, a_2) \} + \{ (\vec{j} \circ a_2) \circ (\vec{K} \circ \mathcal{A}_3) + (\vec{k} \circ a_3) \circ (\vec{J} \circ \mathcal{A}_2) \}/\mathrm{k} \\
\mathcal{B}^2{}_{31} &= \{ \vec{K} \circ (\vec{i} \circ \partial_{w3}\, a_1) + \vec{I} \circ (\vec{k} \circ \partial_{w1}\, a_3) \} + \{ (\vec{k} \circ a_3) \circ (\vec{I} \circ \mathcal{A}_1) + (\vec{i} \circ a_1) \circ (\vec{K} \circ \mathcal{A}_3) \}/\mathrm{k} \\
\mathcal{B}^2{}_{12} &= \{ \vec{I} \circ (\vec{j} \circ \partial_{w1}\, a_2) + \vec{J} \circ (\vec{i} \circ \partial_{w2}\, a_1) \} + \{ (\vec{i} \circ a_1) \circ (\vec{J} \circ \mathcal{A}_2) + (\vec{j} \circ a_2) \circ (\vec{I} \circ \mathcal{A}_1) \}/\mathrm{k} \\
\mathcal{B}^3{}_{00} &= \{ \vec{e} \circ \partial_{s0}\, \mathcal{A}_0 + \vec{i} \circ (\vec{I} \circ \partial_{s1}\, \mathcal{A}_1) + \vec{j} \circ (\vec{J} \circ \partial_{s2}\, \mathcal{A}_2) + \vec{k} \circ (\vec{K} \circ \partial_{s3}\, \mathcal{A}_3) \} \\
&\quad + \{ (\vec{e} \circ \mathcal{A}_0) \circ a_0 + (\vec{I} \circ \mathcal{A}_1) \circ (\vec{i} \circ a_1) + (\vec{J} \circ \mathcal{A}_2) \circ (\vec{j} \circ a_2) + (\vec{K} \circ \mathcal{A}_3) \circ (\vec{k} \circ a_3) \}/\mathrm{k} \\
\mathcal{B}^3{}_{01} &= \{ \vec{I} \circ \partial_{s0}\, \mathcal{A}_1 + \vec{i} \circ (\vec{e} \circ \partial_{s1}\, \mathcal{A}_0) \} + \{ (\vec{e} \circ \mathcal{A}_0) \circ (\vec{i} \circ a_1) + (\vec{I} \circ \mathcal{A}_1) \circ a_0 \}/\mathrm{k} \\
\mathcal{B}^3{}_{02} &= \{ \vec{J} \circ \partial_{s0}\, \mathcal{A}_2 + \vec{j} \circ (\vec{e} \circ \partial_{s2}\, \mathcal{A}_0) \} + \{ (\vec{e} \circ \mathcal{A}_0) \circ (\vec{j} \circ a_2) + (\vec{J} \circ \mathcal{A}_2) \circ a_0 \}/\mathrm{k} \\
\mathcal{B}^3{}_{03} &= \{ \vec{K} \circ \partial_{s0}\, \mathcal{A}_3 + \vec{k} \circ (\vec{e} \circ \partial_{s3}\, \mathcal{A}_0) \} + \{ (\vec{e} \circ \mathcal{A}_0) \circ (\vec{k} \circ a_3) + (\vec{K} \circ \mathcal{A}_3) \circ a_0 \}/\mathrm{k} \\
\mathcal{B}^3{}_{23} &= \{ \vec{j} \circ (\vec{K} \circ \partial_{s2}\, \mathcal{A}_3) + \vec{k} \circ (\vec{J} \circ \partial_{s3}\, \mathcal{A}_2) \} + \{ (\vec{J} \circ \mathcal{A}_2) \circ (\vec{k} \circ a_3) + (\vec{K} \circ \mathcal{A}_3) \circ (\vec{j} \circ a_2) \}/\mathrm{k} \\
\mathcal{B}^3{}_{31} &= \{ \vec{k} \circ (\vec{I} \circ \partial_{s3}\, \mathcal{A}_1) + \vec{i} \circ (\vec{K} \circ \partial_{s1}\, \mathcal{A}_3) \} + \{ (\vec{K} \circ \mathcal{A}_3) \circ (\vec{i} \circ a_1) + (\vec{I} \circ \mathcal{A}_1) \circ (\vec{k} \circ a_3) \}/\mathrm{k} \\
\mathcal{B}^3{}_{12} &= \{ \vec{i} \circ (\vec{J} \circ \partial_{s1}\, \mathcal{A}_2) + \vec{j} \circ (\vec{I} \circ \partial_{s2}\, \mathcal{A}_1) \} + \{ (\vec{I} \circ \mathcal{A}_1) \circ (\vec{j} \circ a_2) + (\vec{J} \circ \mathcal{A}_2) \circ (\vec{i} \circ a_1) \}/\mathrm{k} \\
\mathcal{B}^4{}_{00} &= \{ -\partial_{w0}\, \mathcal{A}_0 - \partial_{w1}\, \mathcal{A}_1 - \partial_{w2}\, \mathcal{A}_2 - \partial_{w3}\, \mathcal{A}_3 \} + \{ (\vec{e} \circ \mathcal{A}_0) \circ (\vec{e} \circ \mathcal{A}_0) \\
&\quad + (\vec{I} \circ \mathcal{A}_1) \circ (\vec{I} \circ \mathcal{A}_1) + (\vec{J} \circ \mathcal{A}_2) \circ (\vec{J} \circ \mathcal{A}_2) + (\vec{K} \circ \mathcal{A}_3) \circ (\vec{K} \circ \mathcal{A}_3) \}/\mathrm{k}
\end{aligned}$$

$\mathcal{B}^4{}_{01} = \{ \vec{e} \circ ( \vec{I} \circ \partial_{w0} \mathcal{A}_1) + \vec{I} \circ ( \vec{e} \circ \partial_{w1} \mathcal{A}_0)\}$
$\quad + \{( \vec{e} \circ \mathcal{A}_0) \circ ( \vec{I} \circ \mathcal{A}_1) + ( \vec{I} \circ \mathcal{A}_1) \circ ( \vec{e} \circ \mathcal{A}_0 )\} /\mathrm{k}$

$\mathcal{B}^4{}_{02} = \{ \vec{e} \circ ( \vec{J} \circ \partial_{w0} \mathcal{A}_2 ) + \vec{J} \circ ( \vec{e} \circ \partial_{w2} \mathcal{A}_0 )\}$
$\quad + \{( \vec{e} \circ \mathcal{A}_0 ) \circ ( \vec{J} \circ \mathcal{A}_2 ) + ( \vec{J} \circ \mathcal{A}_2 ) \circ ( \vec{e} \circ \mathcal{A}_0 )\} /\mathrm{k}$

$\mathcal{B}^4{}_{03} = \{ \vec{e} \circ ( \vec{K} \circ \partial_{w0} \mathcal{A}_3 ) + \vec{K} \circ ( \vec{e} \circ \partial_{w3} \mathcal{A}_0 )\}$
$\quad + \{( \vec{e} \circ \mathcal{A}_0 ) \circ ( \vec{K} \circ \mathcal{A}_3) + ( \vec{K} \circ \mathcal{A}_3 ) \circ ( \vec{e} \circ \mathcal{A}_0 )\} /\mathrm{k}$

$\mathcal{B}^4{}_{23} = \{ \vec{J} \circ ( \vec{K} \circ \partial_{w2} \mathcal{A}_3 ) + \vec{K} \circ ( \vec{J} \circ \partial_{w3} \mathcal{A}_2 ) \}$
$\quad + \{( \vec{J} \circ \mathcal{A}_2 ) \circ ( \vec{K} \circ \mathcal{A}_3 ) + ( \vec{K} \circ \mathcal{A}_3 ) \circ ( \vec{J} \circ \mathcal{A}_2 )\} /\mathrm{k}$

$\mathcal{B}^4{}_{31} = \{ \vec{K} \circ ( \vec{I} \circ \partial_{w3} \mathcal{A}_1 ) + \vec{I} \circ ( \vec{K} \circ \partial_{w1} \mathcal{A}_3 )\}$
$\quad + \{( \vec{K} \circ \mathcal{A}_3 ) \circ ( \vec{I} \circ \mathcal{A}_1 ) + ( \vec{I} \circ \mathcal{A}_1 ) \circ ( \vec{K} \circ \mathcal{A}_3 )\} /\mathrm{k}$

$\mathcal{B}^4{}_{12} = \{ \vec{I} \circ ( \vec{J} \circ \partial_{w1} \mathcal{A}_2 ) + \vec{J} \circ ( \vec{I} \circ \partial_{w2} \mathcal{A}_1 ) \}$
$\quad + \{( \vec{I} \circ \mathcal{A}_1 ) \circ ( \vec{J} \circ \mathcal{A}_2 ) + ( \vec{J} \circ \mathcal{A}_2 ) \circ ( \vec{I} \circ \mathcal{A}_1 )\} /\mathrm{k}$

When the field potential components are limited to the components ($a_0$ , $a_1$ , $a_2$ , $a_3$ ), the following equations can be reduced to the Yang-Mills equations. [3-5]

$\mathcal{B}^1{}_{00} = \{\partial_{s0}\, a_0 - \partial_{s1}\, a_1 - \partial_{s2}\, a_2 - \partial_{s3}\, a_3 \}$
$\quad + \{ a_0 \circ a_0 + ( \vec{i} \circ a_1 ) \circ ( \vec{i} \circ a_1 ) + ( \vec{j} \circ a_2 ) \circ ( \vec{j} \circ a_2 ) + ( \vec{k} \circ a_3 ) \circ ( \vec{k} \circ a_3 )\}/\mathrm{k}$

$\mathcal{B}^1{}_{01} = \vec{i} \circ (\partial_{s0}\, a_1 + \partial_{s1}\, a_0) + \{a_0 \circ ( \vec{i} \circ a_1 ) + ( \vec{i} \circ a_1 ) \circ a_0 \}/\mathrm{k}$

$\mathcal{B}^1{}_{02} = \vec{j} \circ (\partial_{s0}\, a_2 + \partial_{s2}\, a_0) + \{ a_0 \circ ( \vec{j} \circ a_2 ) + ( \vec{j} \circ a_2 ) \circ a_0 \}/\mathrm{k}$

$\mathcal{B}^1{}_{03} = \vec{k} \circ (\partial_{s0}\, a_3 + \partial_{s3}\, a_0) + \{ a_0 \circ ( \vec{k} \circ a_3 ) + ( \vec{k} \circ a_3 ) \circ a_0 \}/\mathrm{k}$

$\mathcal{B}^1{}_{23} = \{ \vec{j} \circ ( \vec{k} \circ \partial_{s2}\, a_3) + \vec{k} \circ ( \vec{j} \circ \partial_{s3}\, a_2)\} + \{( \vec{j} \circ a_2) \circ ( \vec{k} \circ a_3) + ( \vec{k} \circ a_3) \circ ( \vec{j} \circ a_2)\}/\mathrm{k}$

$\mathcal{B}^1{}_{31} = \{ \vec{k} \circ ( \vec{i} \circ \partial_{s3}\, a_1 ) + \vec{i} \circ ( \vec{k} \circ \partial_{s1}\, a_3 )\} + \{( \vec{k} \circ a_3 ) \circ ( \vec{i} \circ a_1 ) + ( \vec{i} \circ a_1 ) \circ ( \vec{k} \circ a_3 )\}/\mathrm{k}$

$\mathcal{B}^1{}_{12} = \{ \vec{i} \circ ( \vec{j} \circ \partial_{s1}\, a_2 ) + \vec{j} \circ ( \vec{i} \circ \partial_{s2}\, a_1 )\} + \{( \vec{i} \circ a_1 ) \circ ( \vec{j} \circ a_2 ) + ( \vec{j} \circ a_2 ) \circ ( \vec{i} \circ a_1 )\}/\mathrm{k}$

So the Yang-Mills equation is the coordinate transformation of the Eq.(6) in strong-weak field. And the above equations show that the Yang-Mills equation is only one special part of definition equation (6).

Table 8. Equations of Yang-Mills field

| | |
|---|---|
| Field potential | $\mathcal{A} = \Diamond^* \circ \mathcal{X}$ |
| Field strength | $\mathcal{B} = (\mathcal{A}/\mathrm{k} + \Diamond) \circ \mathcal{A}$ |
| Field source | $\mu \mathcal{S} = (\mathcal{A}/\mathrm{k} + \Diamond)^* \circ \mathcal{B}$ |
| | $0 = \Diamond^* \circ \mathcal{B}$ |

## 5.2 Three kinds of colors

In the octonionic strong-weak field, the each flavor of lepton particle is supposed to possess the W weak-charge. So the lepton has weak interactions but strong interactions. And there exist two new kinds of unknown field source charges in the nature. It is obvious that the W strong-charge and the S weak-charge are different to the W weak-charge in the strong-weak field, though their field strengths are very close.

If we mix up the W strong-charge with the S weak-charge and W weak-charge in the field theory, it needs three sorts of 'colors' to distinguish the 'mixed' particles. So the each flavor of quark has three kinds of colors in familiar strong interaction theory. [6-8]

Table 9. Quarks comparison of two field theories

| Octonionic strong-weak field | Familiar strong and weak theory |
|---|---|
| quark, (S strong-charge, S weak-charge) | Red quark |
| quark, (S strong-charge, W weak-charge) | blue quark |
| quark, (S strong-charge, W strong-charge) | green quark |

In strong-weak field, each flavor of quark particle is supposed to possess S strong-charge together with other general charge (W weak-charge, W strong-charge or S weak-charge). And the last three general charges are equivalent approximately in the field strength.

In the familiar strong field theory, above three sorts of quarks are believed to the same for each flavor of quark. Therefore it needs three kinds of colors in the familiar quark theory to distinguish those quarks. And in the familiar electro-weak field theory, the three kinds of colors imply that there 'exists' one kind of weaker field with three kinds of colors.

5.3 Quantization of field source particles

5.3.1 Dirac and Schrodinger equations

In the octonionic space, the strong-strong and weak-weak subfields are generated by the particle M which owns W weak-charge and S strong-charge. The S strong-charge current and W weak-charge current of the field source particle m are ($s_0^s$ , $s_1^s$ , $s_2^s$ , $s_3^s$) and ($S_0^w$ , $S_1^w$ , $S_2^w$ , $S_3^w$) respectively. When $\mathcal{U} = 0$ , the wave equation of particle m which moves around M is

$$(\mathcal{W}/c + b\lozenge)^* \circ (\mathcal{M}/b) = 0 \qquad (18)$$

Because of $| s_0^s | \gg | s_i^s |$ and $| S_0^w | \gg | S_i^w |$, then (i = 1, 2, 3)

$$\mathcal{W} = (\mathcal{A}/k + \lozenge)^* \circ \{\mathcal{S} \circ (\mathcal{R} + k_{rx} \mathcal{X})\}$$
$$\approx (k_{sw}\, \mu_w^w\, S_0^w\, \vec{e} \circ \mathcal{V} + k_{rx}\, \mu_s^s\, s_0^s \mathcal{A})/\mu$$
$$+ [\, s_0^s\, \mu_s^s\, \mathcal{V} + k_{sw}\, \mu_w^w\, k_{rx}\, S_0^w\, \vec{e} \circ \mathcal{A}\,]/\mu + \mathcal{A}^* \circ \mathcal{M}/k \qquad (19)$$

When the sum of last two items is equal approximately to 0 and $k_{rx}\, \mu_s^s/\mu = k_{sw}\, \mu_w^w/\mu = 1$, the above equation can be written as

$$\mathcal{W}/c = (sa_0 + wV_0) + \vec{i}\ (sa_1 + wV_1) + \vec{j}\ (sa_2 + wV_2) + \vec{k}\ (sa_3 + wV_3)$$
$$+ \vec{e}\ (k_{sw}\, sA_0 - wv_0) + \vec{I}\ (k_{sw}\, sA_1 - wv_1) + \vec{J}(k_{sw}\, sA_2 - wv_2) + \vec{K}\ (k_{sw}\, sA_3 - wv_3)$$
$$= (p_0 + \vec{i}\ p_1 + \vec{j}\ p_2 + \vec{k}\ p_3 + \vec{e}\ P_0 + \vec{I}\ P_1 + \vec{J} P_2 + \vec{K}\ P_3)$$

where, $S_0^w = wc$ , $s_0^s = sc$ ; $p_j = w_j/c = sa_j + wV_j$ , $P_j = W_j/c = k_{sw}\, sA_j - wv_j$ .

Then

$$0 = (\mathcal{W}/c + b\lozenge)^* \circ \{(\mathcal{W}/c + b\lozenge)^* \circ \Psi\}$$
$$\approx \{(p_0 + b\partial_{s0})^2 - (p_1 + b\partial_{s1})^2 - (p_2 + b\partial_{s2})^2 - (p_3 + b\partial_{s3})^2 - (P_0 + b\partial_{w0})^2 - (P_1 + b\partial_{w1})^2$$
$$- (P_2 + b\partial_{w2})^2 - (P_3 + b\partial_{w3})^2 + bs\lozenge^* \circ \mathcal{A}^* - bw\lozenge^* \circ (\vec{e} \circ \mathcal{V})^*\} \circ \Psi$$

where, $\Psi = - \mathcal{M} \circ \mathcal{I}/b$ is the wave function.

In the above equation, the conservation of wave function is influenced by the field potential, field strength, field source, velocity, S strong-charge and W weak-charge etc. When the weak-weak subfield is equal approximately to 0, and the wave function is $\Psi = \psi(r)\exp(-\mathcal{I}\, Et/b)$, the above equation can be simplified as

$$0 = \{(p_0 - \mathcal{I}\ E/c)^2 - p_1^2 - p_2^2 - p_3^2 - (wv_0)^2 - (wv_1)^2 - (wv_2)^2 - (wv_3)^2$$

$$+ bs\Diamond^{*}\circ\mathcal{A}^{*} - bw\Diamond^{*}\circ(\vec{e}\circ\mathcal{V})^{*}\}\circ\psi(r)$$

$$\approx [(sa_0 - \mathit{I}\,E/c) - (1/2wc)\{p_1^2 + p_2^2 + p_3^2\} + (sb/2wc)\{\Diamond^{*}\circ\mathcal{A}^{*} - \Diamond^{*}\circ(\vec{e}\circ\mathcal{V})^{*}\}]\circ\psi(r) \qquad (20)$$

where, E is the energy, $\mathit{I}$ is an octonionic unit, $\mathit{I}^{*}\circ\mathit{I} = \mathbf{1}$ ; $(sb/2w)(\Diamond^{*}\circ\mathcal{A}^{*})$ is the interplay item of strong-strong subfield with S strong-charge spin.

Limited within certain conditions, Eq.(12) of the quantum mechanics of electromagnetic-gravitational field in the octonionic space can draw some conclusions which are consistent with the Dirac and Schrodinger equations, including the spin and the strong 'moment' etc.

5.3.2 Klein-Gordon equation

In the octonionic space, the strong-strong and weak-weak subfields are generated by the particle M which owns W weak-charge and S strong-charge. The S strong-charge current and W weak-charge current of the field source particle m are $(s_0^s , s_1^s , s_2^s , s_3^s)$ and $(S_0^w , S_1^w , S_2^w , S_3^w)$ respectively. When $\mathcal{L} = 0$ , the wave equation of particle m which moves around M is

$$(\mathcal{W}/c + b\Diamond)\circ(\mathcal{U}/b) = 0 \qquad (21)$$

The above equation can be simplified as

$$(\mathcal{W}/c + b\Diamond)\circ\{(\mathcal{W}/c + b\Diamond)^{*}\circ(\mathcal{M}/b)\} = 0$$

or

$$\{(w_0/c + b\partial_{s0})^2 - (w_1/c + b\partial_{s1})^2 - (w_2/c + b\partial_{s2})^2 - (w_3/c + b\partial_{s3})^2 - (W_0/c + b\partial_{w0})^2 - (W_1/c + b\partial_{w1})^2 - (W_2/c + b\partial_{w2})^2 - (W_3/c + b\partial_{w3})^2\}\circ(\mathcal{M}/b) = 0 \qquad (22)$$

Because of $|s_0^s| \gg |s_i^s|$ and $|S_0^w| \gg |S_i^w|$, then

$$\mathcal{W} = (\mathcal{A}/k + \Diamond)^{*}\circ\{\mathcal{S}\circ(\mathcal{R} + k_{rx}\mathcal{X})\}$$

$$\approx (k_{sw}\,\mu_w^w\,S_0^w\,\vec{e}\circ\mathcal{V} + k_{rx}\,\mu_s^s\,s_0^s\,\mathcal{A})/\mu + [s_0^s\,\mu_s^s\,\mathcal{V} + k_{sw}\,\mu_w^w\,k_{rx}\,S_0^w\,\vec{e}\circ\mathcal{A}]/\mu + \mathcal{A}^{*}\circ\mathcal{M}/k \qquad (23)$$

When the sum of last two items is equal approximately to 0 and $k_{rx}\,\mu_s^s/\mu = k_{sw}\,\mu_w^w/\mu = 1$, the above equation can be written as

$$\mathcal{W}/c = (sa_0 + wV_0) + \vec{i}\,(sa_1 + wV_1) + \vec{j}\,(sa_2 + wV_2) + \vec{k}\,(sa_3 + wV_3)$$
$$+ \vec{e}\,(k_{sw}\,sA_0 - wv_0) + \vec{I}\,(k_{sw}\,sA_1 - wv_1) + \vec{J}(k_{sw}\,sA_2 - wv_2) + \vec{K}\,(k_{sw}\,sA_3 - wv_3)$$
$$= (p_0 + \vec{i}\,p_1 + \vec{j}\,p_2 + \vec{k}\,p_3 + \vec{e}\,P_0 + \vec{I}\,P_1 + \vec{J}P_2 + \vec{K}\,P_3)$$

where, $S_0^w = wc$ , $s_0^s = sc$ ; $p_j = w_j/c = sa_j + wV_j$ , $P_j = W_j/c = k_{sw}\,sA_j - wv_j$ .

Therefore

$$[(p_0 + b\partial_{s0})^2 + (P_0 + b\partial_{w0})^2 + b^2\{\Diamond^2 - (\partial_{s0})^2 - (\partial_{w0})^2\}]\circ(\mathcal{M}/b) = 0 \qquad (24)$$

When the wave function $(\mathcal{M}/b)$ relates only to the quaternionic space, the above equation can be simplified further as

$$\{(wc)^2 + wcb\partial_{s0} + b^2\Diamond_s^2\}\circ(\mathcal{M}/b) = 0 \qquad (25)$$

Limited within certain conditions, Eq.(13) of the strong-weak field in the octonionic space can draw some conclusions which are consistent with Klein-Gordon equation. [9-11]

5.4 The first type of intermediate particles

5.4.1 $\mathcal{G}$ approximate equations

In the octonionic space, the strong-strong and weak-weak subfields are generated by the particle M which owns W weak-charge and S strong-charge. The S strong-charge current and W weak-charge current of the field source particle m are ($s_0^s$ , $s_1^s$ , $s_2^s$ , $s_3^s$) and ($S_0^w$ , $S_1^w$ , $S_2^w$ , $S_3^w$) respectively. When $\mathcal{G} = 0$ , the wave equation of particle m which moves around M is

$$(\mathcal{W}/c + b\Diamond) \circ (\mathcal{A}/b) = 0 \qquad (26)$$

Because of $| s_0^s | \gg | s_i^s |$ and $| S_0^w | \gg | S_i^w |$, then

$$\mathcal{W} = (\mathcal{A}/k + \Diamond)^* \circ \{\mathcal{S} \circ (\mathcal{R} + k_{rx} \mathcal{X})\}$$
$$\approx (k_{sw}\, \mu_w^w\, S_0^w\, \vec{e} \circ \mathcal{V} + k_{rx}\, \mu_s^s\, s_0^s \mathcal{A})/\mu$$
$$+ [\, s_0^s\, \mu_s^s\, \mathcal{V} + k_{sw}\, \mu_w^w\, k_{rx}\, S_0^w\, \vec{e} \circ \mathcal{A}\, ]/\, \mu + \mathcal{A}^* \circ \mathcal{M}/k \qquad (27)$$

When the sum of last two items is equal approximately to 0 and $k_{rx}\ \mu_s^s/\mu = k_{sw}\, \mu_w^w/\mu = 1$, the above equation can be written as

$$\mathcal{W}/c = (sa_0 + wV_0) + \vec{i}\ (sa_1 + wV_1) + \vec{j}\ (sa_2 + wV_2) + \vec{k}\ (sa_3 + wV_3)$$
$$+ \vec{e}\ (\, k_{sw}\, sA_0 - wv_0) + \vec{I}\ (\, k_{sw}\, sA_1 - wv_1) + \vec{J}(\, k_{sw}\, sA_2 - wv_2) + \vec{K}\ (\, k_{sw}\, sA_3 - wv_3)$$
$$= (p_0 + \vec{i}\ p_1 + \vec{j}\ p_2 + \vec{k}\ p_3 + \vec{e}\ P_0 + \vec{I}\ P_1 + \vec{J} P_2 + \vec{K}\ P_3\ )$$

where, $S_0^w = wc$ , $s_0^s = sc$ ; $p_j = w_j/c = sa_j + wV_j$ , $P_j = W_j/c = k_{sw}\, sA_j - wv_j$ .

Then

$$0 = (\mathcal{W}/c + b\Diamond) \circ \{(\mathcal{W}/c + b\Diamond)^* \circ \Psi\ \}$$
$$\approx \{(p_0 + b\partial_{s0})^2 - (p_1 + b\partial_{s1})^2 - (p_2 + b\partial_{s2})^2 - (p_3 + b\partial_{s3})^2$$
$$- (P_0 + b\partial_{w0})^2 - (P_1 + b\partial_{w1})^2 - (P_2 + b\partial_{w2})^2 - (P_3 + b\partial_{w3})^2\ \} \circ \Psi \qquad (28)$$

where, $\Psi = \mathcal{A}/b$ is the wave function.

The above equation can be used to describe the quantum characteristics of intermediate particles which possess spin, S strong-charge and W weak-charge. Limited within certain conditions, Eq.(14) of the quantum mechanics of strong-weak field in the octonionic space can deduce the wave equation and its correlative conclusions.

### 5.4.2 The first Dirac-like equation

In the octonionic space, the strong-strong and weak-weak subfields are generated by the particle M which owns W weak-charge and S strong-charge. The S strong-charge current and W weak-charge current of the field source particle m are ($s_0^s$ , $s_1^s$ , $s_2^s$ , $s_3^s$) and ($S_0^w$ , $S_1^w$ , $S_2^w$ , $S_3^w$) respectively. When $\mathcal{G} = 0$ , the wave equation of particle m which moves around M is

$$(\mathcal{W}/c + b\Diamond) \circ (\mathcal{A}/b) = 0$$

When the energy $\mathcal{W} = 0$, the first Dirac-like equation can be attained from above

$$b\Diamond \circ (\mathcal{A}/b) = 0 \qquad (29)$$

As we know, the Dirac equation can conclude that field source particles (quark and lepton etc.) possess ‘spin’ 1/2. In the same way, the above equation can infer that intermediate particles (intermediate boson etc.) own spin 1 and have no S strong-charge and W weak-charge. And the quantum theory of strong-strong subfield can be obtained.

Limited within certain conditions, Eq.(14) of the quantum theory of the strong-weak field in octonionic space can draw some conclusions which are consistent with the first Dirac-like equation, including the spin of intermediate boson etc. [12-15]

## 5.5 The second type of intermediate particles

### 5.5.1 $\mathcal{T}$ approximate equations

In the octonionic space, the strong-strong and weak-weak subfields are generated by the particle M which owns W weak-charge and S strong-charge. The S strong-charge current and W weak-charge current of the field source particle m are ($s_0^s$ , $s_1^s$ , $s_2^s$ , $s_3^s$) and ($S_0^w$ , $S_1^w$ , $S_2^w$ , $S_3^w$) respectively. When $\mathcal{T} = 0$ , the wave equation of particle m which moves around M is

$$(\mathcal{W}/c + b\diamondsuit)^* \circ (\mathcal{G}/b) = 0 \tag{30}$$

Because of $| s_0^s | \gg | s_i^s |$ and $| S_0^w | \gg | S_i^w |$, then

$$\mathcal{W} = (\mathcal{A}/k + \diamondsuit)^* \circ \{\mathcal{S} \circ (\mathcal{R} + k_{rx}\mathcal{X})\}$$
$$\approx (k_{sw}\,\mu_w^w\, S_0^w \vec{e} \circ \mathcal{V} + k_{rx}\,\mu_s^s\, s_0^s \mathcal{A})/\mu$$
$$+ [\, s_0^s\, \mu_s^s\, \mathcal{V} + k_{sw}\,\mu_w^w\, k_{rx}\, S_0^w \vec{e} \circ \mathcal{A}\,]/\,\mu + \mathcal{A}^* \circ \mathcal{M}/k \tag{31}$$

When the sum of last two items is equal approximately to 0 and $k_{rx}\ \mu_s^s/\mu = k_{sw}\,\mu_w^w/\mu = 1$, the above equation can be written as

$$\mathcal{W}/c = (sa_0 + wV_0) + \vec{i}\ (sa_1 + wV_1) + \vec{j}\ (sa_2 + wV_2) + \vec{k}\ (sa_3 + wV_3)$$
$$+ \vec{e}\ (\, k_{sw}\, sA_0 - wv_0) + \vec{I}\ (\, k_{sw}\, sA_1 - wv_1) + \vec{J}(\, k_{sw}\, sA_2 - wv_2) + \vec{K}\ (\, k_{sw}\, sA_3 - wv_3)$$
$$= (p_0 + \vec{i}\ p_1 + \vec{j}\ p_2 + \vec{k}\ p_3 + \vec{e}\ P_0 + \vec{I}\ P_1 + \vec{J} P_2 + \vec{K}\ P_3\ )$$

where, $S_0^w = wc$ , $s_0^s = sc$ ; $p_j = w_j/c = sa_j + wV_j$ , $P_j = W_j/c = k_{sw}\, sA_j - wv_j$ .

Then

$$0 = (\mathcal{W}/c + b\diamondsuit)^* \circ \{(\mathcal{W}/c + b\diamondsuit)^* \circ \Psi\ \}$$
$$\approx \{(p_0+b\partial_{s0})^2 - (p_1+b\partial_{s1})^2 - (p_2+b\partial_{s2})^2 - (p_3+b\partial_{s3})^2 - (P_0+b\partial_{w0})^2 - (P_1+b\partial_{w1})^2$$
$$- (P_2+b\partial_{w2})^2 - (P_3+b\partial_{w3})^2 + bs\diamondsuit^* \circ \mathcal{A}^* - bw\diamondsuit^* \circ (\, \vec{e} \circ \mathcal{V})^*\ \} \circ \Psi \tag{32}$$

where, $\Psi = \mathcal{G}/b$ is the wave function.

The above equation can be used to describe the quantum characteristics of intermediate particles which possess spin, S strong-charge and W weak-charge. Limited within certain conditions, Eq.(15) of the quantum mechanics of strong-weak field in the octonionic space can deduce the wave equation and its conclusions.

### 5.5.2 The second Dirac-like equation

In the octonionic space, the strong-strong and weak-weak subfields are generated by the particle M which owns W weak-charge and S strong-charge. The S strong-charge current and W weak-charge current of the field source particle m are ($s_0^s$ , $s_1^s$ , $s_2^s$ , $s_3^s$) and ($S_0^w$ , $S_1^w$ , $S_2^w$ , $S_3^w$) respectively. When $\mathcal{T} = 0$ , the wave equation of particle m which moves around M is

$$(\mathcal{W}/c + b\diamondsuit)^* \circ (\mathcal{B}/b) = 0$$

When the energy $\mathcal{W} = 0$, the second Dirac-like equation can be attained from above

$$b\diamondsuit^* \circ (\mathcal{B}/b) = 0 \tag{33}$$

The above equation can infer that intermediate particles own spin N and have no S strong-charge and W weak- charge. Limited within certain conditions, Eq.(15) of strong-weak field in the octonionic space can draw some conclusions which are consistent with the second Dirac-like equation, including the spin of intermediate boson etc.

## 5.6 The third type of intermediate particles

### 5.6.1 $\mathcal{O}$ approximate equations

In the octonionic space, the strong-strong and weak-weak subfields are generated by the particle M which owns W weak-charge and S strong-charge. The S strong-charge current and W weak-charge current of the field source particle m are ($s_0^s$ , $s_1^s$ , $s_2^s$ , $s_3^s$) and ($S_0^w$ , $S_1^w$ , $S_2^w$ ,

$S_3^w$) respectively. When $\mathcal{O} = 0$ , the wave equation of particle m which moves around M is

$$(\mathcal{W}/c + b\lozenge) \circ (\mathcal{T}/b) = 0 \qquad (34)$$

The above equation can be simplified further as

$$(\mathcal{W}/c + b\lozenge) \circ \{(\mathcal{W}/c + b\lozenge)^* \circ (\mathcal{G}/b)\} = 0$$

or

$$\{(w_0/c + b\partial_{s0})^2 - (w_1/c + b\partial_{s1})^2 - (w_2/c + b\partial_{s2})^2 - (w_3/c + b\partial_{s3})^2 - (W_0/c + b\partial_{w0})^2 - (W_1/c + b\partial_{w1})^2 - (W_2/c + b\partial_{w2})^2 - (W_3/c + b\partial_{w3})^2\} \circ (\mathcal{G}/b) = 0 \qquad (35)$$

Because of $| s_0^s | \gg | s_i^s |$ and $| S_0^w | \gg | S_i^w |$, then

$$\mathcal{W} = (\mathcal{A}/k + \lozenge)^* \circ \{\mathcal{S} \circ (\mathcal{R} + k_{rx}\mathcal{X})\}$$
$$\approx (k_{sw}\, \mu_w^w\, S_0^w\, \vec{e} \circ \mathcal{V} + k_{rx}\, \mu_s^s\, s_0^s \mathcal{A})/\mu$$
$$+ [\, s_0^s\, \mu_s^s\, \mathcal{V} + k_{sw}\, \mu_w^w\, k_{rx}\, S_0^w\, \vec{e} \circ \mathcal{A}\,]/\, \mu + \mathcal{A}^* \circ \mathcal{M}/k \qquad (36)$$

When the sum of last two items is equal approximately to 0 and $k_{rx}\, \mu_s^s/\mu = k_{sw}\, \mu_w^w/\mu = 1$, the above equation can be written as

$$\mathcal{W}/c = (sa_0 + wV_0) + \vec{i}\, (sa_1 + wV_1) + \vec{j}\, (sa_2 + wV_2) + \vec{k}\, (sa_3 + wV_3)$$
$$+ \vec{e}\, (\, k_{sw}\, sA_0 - wv_0) + \vec{I}\, (\, k_{sw}\, sA_1 - wv_1)$$
$$+ \vec{J}(\, k_{sw}\, sA_2 - wv_2) + \vec{K}\, (\, k_{sw}\, sA_3 - wv_3)$$
$$= (p_0 + \vec{i}\, p_1 + \vec{j}\, p_2 + \vec{k}\, p_3 + \vec{e}\, P_0 + \vec{I}\, P_1 + \vec{J} P_2 + \vec{K}\, P_3\,)$$

where, $S_0^w = wc$ , $s_0^s = sc$ ; $p_j = w_j/c = sa_j + wV_j$ , $P_j = W_j/c = k_{sw}\, sA_j - wv_j$ .

Therefore

$$[(p_0 + b\partial_{s0})^2 + (P_0 + b\partial_{w0})^2 + b^2\{\lozenge^2 - (\partial_{s0})^2 - (\partial_{w0})^2\}] \circ (\mathcal{G}/b) = 0 \qquad (37)$$

When the wave function ($\mathcal{G}/b$) relates only to the quaternionic space, the above equation can be simplified further as

$$\{(wc)^2 + wcb\partial_{s0} + b^2\lozenge_s^2\} \circ (\mathcal{G}/b) = 0 \qquad (38)$$

The above equation can be used to describe the quantum characteristics of intermediate particles which possess S strong-charge, W weak-charge and spin N. Limited within certain conditions, Eq.(16) of the quantum mechanics of the strong-weak field in the octonionic space can deduce the wave equation and its relative conclusions.

5.6.2 The third Dirac-like equation

In the octonionic space, the strong-strong and weak-weak subfields are generated by the particle M which owns W weak-charge and S strong-charge. The S strong-charge current and W weak-charge current of the field source particle m are ($s_0^s$ , $s_1^s$ , $s_2^s$ , $s_3^s$) and ($S_0^w$ , $S_1^w$ , $S_2^w$ , $S_3^w$) respectively. When $\mathcal{O} = 0$ , the wave equation of particle m which moves around M is

$$(\mathcal{W}/c + b\lozenge) \circ (\mathcal{T}/b) = 0$$

When the energy $\mathcal{W} = 0$, we obtain the third Dirac-like equation

$$b^2\, \lozenge^2 \circ (\mathcal{B}/b) = 0 \qquad (39)$$

The above equation can be used to describe the quantum characteristics of intermediate particles which possess spin N and have no S strong-charge and W weak-charge. Limited within certain conditions, Eq.(16) of strong-weak field in the octonionic space can deduce the third Dirac-like equation and its conclusions.

If the preceding field potential, field strength and field source are extended to four types of field potential, field strength and field source of strong-weak field, many extra and also more complicated equations set can be gained from Eqs.(5) ~ (16) to describe the copious physical characteristics of the strong and weak interactions. [16, 17]

5.7 Compounding particles

In strong-weak field, the intermediate and field source particles can be tabled as following. In Table 10, we can find intermediate boson, lepton, three colors of quarks, and other kinds of new and unknown particles which may be existed in the nature and expressed in parentheses.

Table 10. Sorts of particle in strong-weak field

| | S strong-charge | W strong-charge | S weak-charge | W weak-charge |
|---|---|---|---|---|
| S strong-charge | (intermediate particle) | green quark | red quark | blue quark |
| W strong-charge | green quark | (intermediate particle) | (?) | (?) |
| S weak-charge | red quark | (?) | (intermediate particle) | (?) |
| W weak-charge | blue quark | (?) | (?) | lepton, intermediate boson |

In the electromagnetic-gravitational field, the field sources and intermediate particles can be tabled as following. In the Table 11, we can find the photon, electron, and other kinds of new and unknown particles which may be existed in the nature.

Specially, by analogy with above three colors of quarks in Table 10, it can be predicted that there exist the 'green electron' and 'red electron' in the nature, if the familiar electron is denoted as the 'blue' electron.

Some particles may possess more than two sorts of general charges. If the quarks possess the S strong-charge, W weak-charge, E electric-charge and G gravitational-charge, they will take part in strong, weak, electromagnetic and gravitational interactions.

Table 11. Sorts of particle in electromagnetic-gravitational field

| | E electric-charge | G electric-charge | E gravitational -charge | G gravitational -charge |
|---|---|---|---|---|
| E electric-charge | photon | (green electron, dark matter) | (red electron, dark matter) | (blue) electron |
| G electric-charge | (green electron, dark matter) | (intermediate particle) | (dark matter) | (dark matter) |
| E gravitational -charge | (red electron, dark matter) | (dark matter) | (intermediate particle) | (dark matter) |
| G gravitational -charge | (blue) electron | (dark matter) | (dark matter) | mass, (intermediate particle) |

## 6. Conclusions

By analogy with the octonionic electromagnetic-gravitational field, the octonionic strong-weak field theory has been developed, including the field equation, its quantum equations and some new unknown particles.

In the strong-strong subfield, the study deduces the Yang-Mills equation, Dirac equation, Schrodinger equation and Klein-Gordon equation of the quarks etc. And it is able to infer the three sorts of Dirac-like equations of intermediate particles among quarks. It may predict that there are some new energy parts, new unknown particles of field sources (quarks) and their intermediate particles in the strong-strong subfield.

In the weak-weak subfield, the research infers the Yang-Mills equation, Dirac equation, Schrodinger equation and Klein-Gordon equation of leptons etc. And it is able to conclude the three types of Dirac-like equations of intermediate bosons among leptons. It may predict that there are some new energy components, new unknown particles of field sources (leptons) and their intermediate particles in the weak-weak subfield.

In the strong-weak field, the paper explains why the quarks possess three sorts of colors and take part in four kinds of interactions. The study predicts that there may exist two kinds of unknown subfields, which field strengths may be equivalent to that of the weak-weak subfield. Sometimes, we may mix up mistakenly the first two subfields with weak-weak subfield. So it needs three sorts of 'colors' to distinguish those 'mixed' subfields in the quark theory.

## Acknowledgements

The author thanks Yun Zhu, Zhimin Chen and Minfeng Wang for helpful discussions. This project was supported by National Natural Science Foundation of China under grant number 60677039, Science & Technology Department of Fujian Province of China under grant number 2005HZ1020 and 2006H0092, and Xiamen Science & Technology Bureau of China under grant number 3502Z20055011.